\documentclass [10pt,a4paper]{article}
\usepackage[margin=2.5cm]{geometry}
\usepackage{graphicx} 
\usepackage{todonotes}
\usepackage{float}

\title{A non-custodial wallet for digital currency:\\design challenges and opportunities}
\author{Ryan Bowler, Geoffrey Goodell, Joe Revans, Gabriel Bizama, Chris Speed}
\date{May 2024}

\begin{document}
\maketitle

\section{Abstract}

Digital currency is a novel form of money that could be issued and regulated by central banks or other actors, offering benefits such as programmability, security, and privacy. However, the design of a digital currency system presents numerous technical and social challenges. This article presents the design and prototype of a \textit{non-custodial wallet}, a device that enables users to store and spend digital currency in various contexts. To address the challenges of designing a digital currency system, we conducted a series of workshops with internal and external stakeholders, using methods such as storytelling, metaphors, and provotypes to communicate digital currency concepts, elicit user feedback and critique, and incorporate normative values into the technical design. We derived basic guidelines for designing digital currency systems that balance technical and social aspects, and reflect user needs and values. Our work contributes to the digital currency discourse by demonstrating a practical example of how digital currency could be used in everyday life and by highlighting the importance of a user-centred approach.

\section{Introduction}

Money is deeply embedded in many areas of society, connecting social, economic, and political discourse~\cite{ingham2020money}. It is considered a social technology~\cite{ingham1996money} used by millions of people at any given time.  With the advent of the digital economy and the decreasing use of money in its tangible state of cash~\cite{ward2019understanding} in favour of digital representations stored on physical devices and accessible through software, our understanding of money continually evolves and adapts. The rise of cryptocurrency heralds a significant shift in the popular perception and use of money, with thousands of digital coins being produced for various applications~\cite{Gomzin2022} and acting as unregulated forms of money on trading platforms.

Digital currency provides an opportunity for designers to incorporate properties of established forms such as the anonymity of cash or the digital convenience of debit - things that people value. However, it is still up for debate whether people will use these ‘coins’. Some versions of digital currency may only be used on the wholesale market and not by everyday people. Because of the complexity intrinsic to retail forms of money~\cite{panetta2022demystifying}, infrastructural changes and diverse human perspectives will be crucial in the design of digital currency if it is to be used on a daily basis for retail transactions. The use of digital currency for retail purposes requires an understanding of the diverse needs of potential users. However, the discourse surrounding this digital ‘coin’ often overlooks the user angle~\cite{abramova2022can} and focuses mainly on its technical details~\cite{auer2021cbdc}. All scenarios of digital currency should be well-understood and appropriate methods, technologies, and systems must be in place to store, validate, and operate its usage within varying transactional scenarios. This means the design corpus around this new form of currency will be complex and must encompass technical infrastructure along with user narratives.

Governments and central banks worldwide are looking to harness technologies and concepts similar to those used in cryptocurrency to create a sovereign digital ‘coin’ known as Central Bank Digital Currency (CBDC). Just as central banks have managed the distribution and redemption of money in the form of cash for centuries, CBDC offers the chance to do the same with digital assets. Several countries are currently trialling, testing, or discussing CBDC as a form of money within their economies. This means that millions of people could soon be interacting with these coins as a legitimate and institutionally backed form of currency. The Bank of England and the HM Treasury are some of the latest to engage in CBDC discussions~\cite{UKDigitalPound2023}.  The ongoing debate concerning future CBDC architectures, and the possibility that CBDC or other forms of digital currency might represent the future of money for households and businesses, is an important motivation for investigating the requirements for infrastructure and tools that the general public can use to exchange and use digital currency.

To contribute to the emergent digital currency discourse and provoke interest in user-focussed digital currency design, we explored ways people would come to use this digital ‘coin’ using a series of design methods. This resulted in the creation of our own non-custodial digital currency wallet - a software/hardware device that stores and allows the transaction of digital currency within a retail setting. We ran workshops to understand complex digital currency systems, align our team’s understanding, refine the journey of a digital currency asset, and explore human factors from the perspectives of different users. We used methods such as storytelling, metaphoric language, user journey mapping, and `provotypes' to aid in the creation of a series of non-custodial wallets. Each method had its benefits and downsides. By matching different wallet features, we arrived at our final prototype wallet which we termed ‘The Minimum Standalone Wallet’. In this article, we characterise our approach, the challenges we encountered, and our overall design output.

\section{What is the value proposition of digital currency?}

For 35 years, siblings Simon and Victoria have run a small pub in Leicester, England. Every Sunday, starting at 11am, they welcome many local residents who come to enjoy what is deemed to be the best roast in the East Midlands. For the last two years, the pub only accepts card payments, mainly due to the transaction costs involved when accepting cash. Previously, Simon and Victoria held the cash throughout the month and then brought it to their nearest branch of a large local bank, about two kilometres away. The process to deposit the cash was burdensome on Simon and Victoria and left them responsible for large amounts of cash in the interim, inviting the risk of loss or theft. Digital payments are more convenient for the business and many of its visitors, and digital payments avoid some of the risk of theft or loss associated with cash. However, not all of their previous customers have bank accounts, and the use of cards exposes customers to unwanted profiling. As a result, although Simon and Victoria managed to reduce their costs, implementing their new policy of refusing cash payments involved turning some of their loyal customers away.

Electronic payments earned through card payment networks by many people just like Simon and Victoria could one day be replaced by a digital currency. Simon and Victoria would have both a public payment option for consumers without bank accounts or who wish to pay with assets they own, while retaining the security and comfort of instantly depositing payments received at the pub. In addition, Simon and Victoria would have an efficient mechanism for demonstrating tax compliance, providing receipts to customers, and proving to customers and authorities that money will be received into a valid bank account. It would also allow customers without local bank accounts (e.g. tourists and, increasingly, digital nomads), local merchants, and suppliers to make and accept affordable and efficient payments directly from their digital wallets. High transaction fees would be avoided by merchants and tourists alike. While the small business run by Simon and Victoria is located in England, this scenario exists across the globe: in advanced and emerging economies, in urban and rural areas, in Europe, Asia, Africa, Middle East, and Latin America, where the choice to refuse cash is increasingly palatable, despite the lack of a public payment option. 

A digital currency could positively impact inclusion and the economy by expanding access to and usage of affordable financial services. In response to a decline in the use of cash, a digital currency could be introduced as an additional form of public money and means of payment. In this sense, a digital currency could be inexpensive, easy to use, secure, private by design for consumers, safety-enhancing, and fast. Payment-related issues, such as domestic payments efficiency, consumer privacy, and payments safety, remain a top policy priority for both advanced and emerging economies for issuing a general purpose digital currency. According to data collected by the Bank for International Settlements (BIS) from central banks, there are 28 pilots and 68 central banks have communicated publicly about their CBDC work~\cite{bis2021,bis2022}.  Digital currency could also contribute to social inclusion and social sustainability~\cite{eu2021}, whether or not it takes the form of CBDC. 

A well-designed digital currency for public use would:

\begin{itemize}

\item Expand modes of payments beyond financial institutions

\item Reduce merchant acceptance costs of digital payments

\item Develop an innovative and competitive local ecosystem 

\item Reduce the cost of cross-border payments 

\item Foster interoperability 

\end{itemize}

\subsection{Expand modes of payments beyond financial institutions}

Development of a digital currency could support financial inclusion through the design of low-cost solutions both for consumers and Micro, Small and Medium Entreprises (MSMEs), specifically merchants accepting payments in digital currencies. A digital currency could lead to the development of a real-time payments system, allowing for transactions to settle and finalize in seconds, notwithstanding the day and time of the transaction. 

\subsection{Reduce merchant acceptance costs of digital payments}

One of the main friction points for the adoption of digital payments by merchants is acceptance costs and fees.

Costs remain high for merchants accepting retail payments due to the need for multiple intermediaries to complete a transaction. Globally, the average fees for retailer acceptance amount to approximately 3-5 percent of the total amount sent. Some jurisdictions have managed to decrease fees with the rise of mobile money usage. For example, in Latin America, the cost for merchants, mainly MSMEs, to accept a payment amounts to 1-2.5 percent in transaction fees depending on the payment instrument used by the consumer when purchasing goods. In the past few years, innovation brought more affordable tools for retailers such as QR payments and mobile money. However, fees to accept a digital payment made by a consumer through a debit and/or credit card are still very high for merchants. 

Digital currency could reduce merchants' acceptance costs due to the efficiency to move value and the removal of intermediaries required to process a transaction.  We note that removing intermediaries is not a trivial undertaking.  In particular, fair exchange between two parties is impossible without some kind of trusted third party to the transaction~\cite{pagnia1999}, and in modern retail payments, the role of third party is played by an institutional framework mediated by asset custodians on both sides of the transaction.  To allow users to possess and control digital assets directly requires a new approach.

It might be tempting to use cryptocurrencies, including stablecoins such as USDT, for this task, but cryptocurrencies have important weaknesses.  Most importantly, cryptocurrencies generally rely upon permissionless DLT networks, which feature massive negative externalities, such as susceptibility to denial of service or takeover by hostile parties that easily be regulated, as well as a transaction processing mechanism that is inefficient, expensive, and environmentally destructive.  Most cryptocurrencies are not private by design, and those that are, such as ZCash and Monero, require additional, computationally expensive cryptographic operations and often run afoul of regulations.  Finally, because running a validator node in a cryptocurrency network carries a high fixed cost that typical users cannot afford to bear, most retail consumers rely upon (custodial) wallet operators to manage their interaction with the system and cannot generally expect to control them directly, even if they store their keys on hardware wallets.

\subsection{Develop an innovative and competitive local ecosystem}

Having digital currency could be an opportunity for the financial sector to enhance further innovative payment services beyond traditional services offered by banks.

In a two-tier distribution model for CBDC, central banks would issue a CBDC and remain in control of monetary policy, and financial entities would distribute a CBDC to consumers and MSMEs, minimising the risk of banking disintermediation that affects the cost of credit and financial stability. A well-designed CBDC would also ensure that regulated financial entities can validate and process payments, enabling competition. 

This architecture would in turn incentivise the development of innovative solutions by financial entities in a competitive and secure environment as consumer needs evolve. A clear example of this could be products that facilitate transactions for consumers with limited access to telecommunications services, specifically in rural regions or for vulnerable groups with a low rate of digital and financial literacy.

\subsection{Reduce the cost of cross-border payments}

The global average cost of cross-border payments remains very high at 6.25 percent of the amount sent. This is more than double the United Nations’ Sustainable Development Goal target of 3 percent. 

The high costs of remittance payments have real-world impacts on individuals and businesses. High costs affect, for example, migrants seeking to send money home to their families, MSMEs importing products from foreign suppliers, and nonprofit organizations seeking to deliver relief funds abroad. These high fees result in suboptimal outcomes for economies and societies at large.

A well-designed digital currency could enable cross-border payments including remittances and business-to-business payments that finalize in seconds and cost less than one percent, mainly due to the reduction of intermediation.  It is worth noting that despite the inefficiency and environmental costs of cryptocurrency systems that spend computational resources running smart contract code and defending against takeover by hostile parties, not all digital currency systems must be designed this way.  Permissioned distributed ledger systems, such as the model upon which we base our core proposal~\cite{goodell2021scalable}, do not require a ledger-based model for code execution, a ledger-based model for managing tokens, or a resource-based approach to defending against hostile actors.  In the absence of such features, the cost for operating the system is greatly reduced and, in our estimation, may perform favourably when compared to the cost of operating the existing correspondent banking system, which relies upon costly operational and reconciliation procedures for every transaction.  In particular, the model that we consider features zero marginal cost to the ledger for each transaction, since assets carry their own state in the form of proofs of provenance, validator nodes are responsible for performing aggregation, and the ledger grows at a fixed rate.

The primary focus of this investigation is domestic retail payments, not cross-border payments.  However, we believe that standardisation will enable devices to interoperate with currencies of multiple issuers, which can certainly be multiple national governments.

\subsection{Foster interoperability}

A digital currency could provide instant real-time payments in central bank money, so payees (including merchants) could receive funds instantly. The core ledger could be built with relatively simple functionality, so that it would be as efficient and cost-effective as possible, expanding the offering of financial services. 

An effective design model should prioritise interoperability, permitting that any digital currency user would be able to pay any other digital currency user, without requiring the payer to have or use a custodial account of any kind. These solutions and products can enable a digital currency to meet payment needs as they evolve. A digital currency should also be interoperable with cash and electronic money at a domestic and retail level. 

A well-designed digital currency could bring positive effects to inclusion and the broader economy. It could improve payments efficiency and safety but more importantly could contribute to social inclusion, develop an innovative and competitive local ecosystem, reduce the cost of remittances for migrants, and foster interoperability in the financial sector.

\section{Navigating a Shifting Payments Landscape}

The concept of socio-technical systems allows designers to understand the interconnected workings of society, from individuals to institutions, and how these relate to technical systems~\cite{baxter2011socio}. Money fits into this social technology as it cannot be fully understood without considering the social relationships from which it emerges~\cite{ingham1996money}. A socio-technical system evolves alongside the changing social, cultural, and political needs of the context in which it is designed~\cite{kling1999and}. As society changes, so does the way money is perceived and used.

This is demonstrated by the decrease in the use and acceptance of established forms of money, especially cash. This trend is amplified by multiple factors ranging from social and technological changes~\cite{arvidsson2019building},~\cite{arvidsson2018future}, to the COVID-19 pandemic~\cite{UKFinance2021}, to the emergence of systems for digital payments that offer individual ownership of money and provide greater financial inclusion for specific individuals~\cite{demirguc2018global}. Physical acceptance of cash by retailers has also changed. After the first lockdown uplift, 42 per cent of people in the UK experienced interactions with retailers who no longer accept cash payments~\cite{caswell2020cash}, a trend that continues. Further hampering cash, specifically in the UK, is the fact that the government has not proposed any laws to prohibit the non-acceptance of cash~\cite{govuk_new_powers_to_protect_access_to_cash}.

This becomes potentially a cautionary issue, as designers have shown that cash remains important to many people. This includes older generations who may prefer using cash and have created routines around it~\cite{vines2012eighty}, as well as those for whom the physical cost of using digital forms is prohibitive due to poor design that exacerbates an already present digital divide~\cite{dai2023cognitive,bon2020digital}. Because of these limitations inherent to digital forms of money, no matter how money evolves in the future, physical forms of money like cash are likely to remain pivotal as a tangible asset upon which people rely for their engagement with the economy. For instance, cash possesses many valuable properties. It can be obtained instantaneously, making it preferable over cashless payments that can take hours or even days to process~\cite{kameswaran2019cash}. Cash can also be easily converted into other assets and serves as both a sovereign currency and a store of value. Its distribution is monitored by central banks, and its creation is governed in a way that supports its legitimacy.  Central banks manage cash using a variety of mechanisms, ranging from monitoring commercial banks to literally printing money so that people can purchase goods~\cite{bankofengland}.

However, cash also has flaws that can be addressed by digital forms of money. For example, cash is not suitable for payments at a distance, and it has has size restrictions that limit the amounts that can be transacted at a single time and place. Digital forms of money are sometimes bound by fewer restrictions on its transport and use~\cite{didenko2019evolution}. Cash is visual, making it challenging to use if a person cannot visually denote its value, currency, or amount~\cite{kameswaran2019cash}. Cash can be vulnerable to counterfeiting, physical theft, regulatory evasion, and lack of visibility into its transactions, all of which can support criminal activity. However, it is not clear whether removing banknotes, especially those of high value, might not eliminate crime but only change its form, such as a shift toward an increase in institutional corruption or the use of commodity money, such as precious metals or stones, or criminal activities~\cite{mcandrews2020case}. Limitations inherent to cash have been presented by central banks across the world as a justification for pursuing CBDC. With 38 per cent of global countries either piloting CBDC or expressing public interest in it~\cite{patel2022cbdc}, CBDC has the potential to change our concept of money in unprecedented ways across the globe. Nevertheless, the properties of cash, whether positive or negative, remain valuable to many people.

As digital currency may become more prevalent in the future, designers are exploring ways to ensure continued accessibility to cash, particularly for those who rely on it or to support economic continuity in the event of infrastructural disruptions, such as power outages. A variety of potential designs have been proposed, including those that allow cash-like functionality without cash distribution infrastructure. One proposal is to allow users to self-print cash that has features compatible with digital payment systems, such as taking a picture to turn the cash into digital forms~\cite{zhou2023print}. Consumer affordances of cash, such as anonymity, are becoming important considerations in the design of digital currency~\cite{wust2022platypus}. Yet, any design attempt is also thwarted by the theoretical unknowns of exactly what a digital currency system will come to be. Will it be issued by central banks, or privately, like Meta’s Libra coin, which is underpinned by financial assets~\cite{diez2020cbdc}? Will digital currency be distributed as a token by banks, or will people have accounts with the central bank~\cite{bindseil2019central}? Will digital currency be used in wholesale markets, retail markets, or both? Designing a retail digital currency would be significantly more challenging due to the complexity of factors associated with retail usage~\cite{panetta2022demystifying}. Currently, several banks are working on digital currency as both a cross-border payment system and a consumer-focussed system~\cite{auer2020taking}, making digital currency not only a country-specific narrative but a global one. Though a wide range of topics are covered in digital currency discourse, gaps in research focus have emerged. One example is the majority of focus on technical aspects while user-focussed approaches have been based on theory rather than practical methodologies~\cite{barzykinclusive}.

Researchers have employed design thinking to account for user requirements, for instance, inclusive design~\cite{clarkson2013inclusive} and a suite of methodologies to explore user requirements~\cite{courage2005understanding}. However, designers do not really have a design framework or set of methodologies for approaching design in money narratives~\cite{elsden2022designing}, especially in the context of CBDC, making user-centric design around money and digital currency discourse a relatively novel concept and space of exploration.

In our research, we focus primarily upon digital currency issued by central banks in the form of CBDC, and the assumption that we would be working with an institutionally supported digital currency has formed the basis for our attention to institutional participants, including central banks, in our workshops.  However, whether or not future money takes the form of CBDC, stablecoins, reserve-backed tokens~\cite{goel2023}, or something else, institutionally supported digital currency represents a novel form of money that proponents argue will integrate with established forms of payment, such as cash, even if use of those forms of payment continues to decrease. This suggestion demonstrates that the monetary landscape remains unknown, continues to evolve, and provides an opportunity for designers to play a significant role. As proposed digital currency architectures continue to evolve, it is increasingly important for designers to bridge the research gap and seek to address the many questions that remain, including the value that people place on current forms of money and how those forms of money might fit into a world with digital currency. Designers can also work on formulating frameworks, methods, and user-centric approaches for digital currency systems and designs.

\section{Designing a non-custodial wallet}

\subsection{Challenges in creating a digital currency system}

Many questions persist in the debate about the right set of requirements for digital currency, and equally just as many design opportunities have presented themselves. We contribute to this development by offering our own approach to the design of a digital currency system. This approach took many steps, iterations, and methods. All of which we discuss in the following sections.

Developing a new digital currency system, from conceptualisation to design, presented several unique challenges. One of the main challenges was figuring out how to incorporate normative values such as social, cultural, or political factors into the technical design requirements. When designing for something as complex as money, it’s important to consider it as a social relation. This means that designing its tangible and intangible properties involves more than just the form in which it is presented. It also involves designing the values that people associate with what money does, creates, or upholds~\cite{halloluwa2018value}.

Design researchers have developed methods for designing socio-technical systems in specific contexts~\cite{dardenne1993goal,bryl2009designing,maguire2014socio}. However, their findings do not fully translate to designing for concepts like money and values. This results in a lack of framework and tools for generating shared understandings of values between designers and stakeholders~\cite{pommeranz2012elicitation}. This also means that there is little guidance available, opening up the opportunity for an exploratory design approach to money and digital currency systems. To address this challenge, we formulated our own guidelines for creating value within our digital currency system based on principles such as a need for diverse payment options, individual asset ownership, and private digital payments. These principles were derived from incorporating key features of tangible cash, such as anonymity, security, not requiring online access~\cite{krueger2018pros}, and its fungibility, accessibility, non-discrimination, and direct ownership. This approach seemed appropriate because the value people and society place on the usage of cash are firmly established~\cite{vines2012eighty,kameswaran2019cash,mcandrews2020case}.

A second challenge that arose was addressing the difficulty in understanding digital currency. digital currency is a complex topic and not always accessible for everyday people due to its novelty, its wide range of technical terms, and the complexity of the current payment infrastructures with which it would integrate. This problem of understanding complex technologies is not new and is referred to as ‘technical debt’ in software disciplines~\cite{theodoropoulos2011technical}. To develop a shared understanding and vision for the system among designers and stakeholders who were unfamiliar with digital currency discourse, we employed techniques inspired by design methods that use metaphors to bolster creativity with non-experts~\cite{ylipulli2017creative} and to communicate technical terms through metaphorical analogies in disciplines like computer science~\cite{van1988binary}. Metaphors helped us bridge pre-existing known concepts with the technical systems of digital currency. For example, we used metaphors such as “USO assets are like a sheet of infinitely extensible paper” (as a way to describe an asset that maintains its own state), “blind signatures are like signing an opaque, carbon-lined envelope” (as a way to describe how a signer can recognise a signature as valid without recognising what had been signed), and “CBDC assets are hot when you create them, so you need to wait for them to cool down before spending” (as a way to explain the process of accumulating a sufficiently large anonymity set). These metaphors and allegories were then reused and reworked during further workshop activities with internal and external stakeholders.


The final challenge was to effectively find a way to demonstrate the system’s technical specifications, as well as its benefits and drawbacks, to potential users and stakeholders. Compiling all these challenge points together allowed for the digital currency design work to be separated into three strands:

\begin{itemize}
    \item \textit{meta-design}, which focuses on designing working practices, knowledge exchange, and stakeholder engagement~\cite{bauer2009designing,fischer2000,fischer2004};
    \item \textit{technical systems design}, which focuses on designing how the system should function and be built~\cite{bacon2005,goodman2011,cockton2019}; and
    \item \textit{interaction design}, which focuses on designing how the system interfaces with the wider world~\cite{fischer2007,herrmann2014,thornton2022,schutz2020}.
\end{itemize}

These three strands provide a framework for understanding the integrity of the project by examining how each one interacts within our digital currency system concept. Each strand was not approached in a linear fashion, but rather in different non-linear steps, with each influencing and reacting to the outcomes of the others. The overall binding challenge points, as well as the considerations and choices that emerged, created a starting point for understanding the prerequisites of our digital currency system’s technical requirements. This led us to return to our inspirations of values that emerge from cash as a point of reference, bolstering our original requirements.

\begin{itemize}
    \item \textit{Ownership}: The digital currency system must be token-based (not account-based). The tokens must be government-issued (i.e., an analogue to banknotes and coins) and unforgeable by design~\cite{goodell2021scalable}.
    \item \textit{possession and control}: Users must have the option to store tokens directly, outside the context of accounts, in non-custodial wallets. Non-custodial wallets must not be identifiable, issued by third parties, or registered, and they must not require trusted computing or certified hardware. The system must be private by design for consumers.
    \item \textit{Privacy}: The identity of somebody who sends money must not be linked to the recipient of the transaction, the value of the transaction, and the transaction metadata (e.g., time, location, service providers, etc.).
    \item  \textit{Legal}: The system must be compatible with anti-money laundering requirements for recipients of money, and authorities should have a way to identify the recipient of money, e.g. a vendor of consumer goods, in most transactions. This requirement introduces a limitation to whether or not peer-to-peer transactions are allowed within the system, and generally implies that recipients will not have the same degree of privacy as payers.  The requirement for partial transparency implies that some facilitators of payments, for example a bank that allows vendors to deposit money received from customers, would be subject to regulations and auditable as a means of managing risk.
    \item \textit{Maintaining value}: The system must support the two-tiered banking system, wherein central banks issue money and private-sector banks make risky investments.  We intend for the design to respect the overall structure of the existing financial system, with only the changes that are needed to support digital cash.  The digital currency system should interoperate with today's set of institutions and infrastructure and should not be seen as a wholesale replacement for them. Clearing, settlement, and other operations should be performed by regulated, private sector organisations such as banks and other money services businesses, although the system should be overseen by a central bank, as is commonplace among payment systems operating within many jurisdictions today.
\end{itemize}

These finalised requirements provided guidance on the directions we could take in formulating our digital currency system through our design practice. We began to realise that certain technical components were required within our design output. For example, blind signatures, which provide privacy by design with measurable anonymity, fit our design guidelines. Chaum (1983) suggests that blind signatures could be a method to achieve privacy and prevent criminality in digital payment systems at the same time~\cite{chaum1983blind,chaum2021}. In this approach, the content of a transaction is disguised from the person signing it, allowing transactions to be validated and legality upheld while maintaining the privacy of the payer. Another component could be a public permissioned distributed ledger technology (DLT) system for decentralised transaction processing, with DLT nodes operated by independent payment service providers. A DLT system is like a big shared notebook that keeps track of transactions. Anyone can look at it, but only certain people are allowed to write in it. This means that different people and companies can work together to process transactions without needing a single entity to be in charge. Finally, unforgeable, stateful, oblivious (USO) assets could be used to avoid requiring issuers to keep track of the integrity or ownership of individual assets. A USO asset keeps track of its own history and can demonstrate its own legitimacy, without issuers or financial institutions having to keep track of the coin’s status as it passes from one person to another. Complexity aside, these technologies provide features that support adherence to our guidelines and provide a starting point for envisioning the digital currency system that we would come to design~\cite{goodell2021scalable}.

\subsection{Human Centred Central Bank Digital Currency}

While digital currency design proposals, including ours, have received increasing attention for their technical characteristics, we believe it is crucial to also focus on the interaction aspects of our digital currency system and adopt a human-centred perspective. We created a workshop called “Human-centred digital currency” based on the Socio-Technical Walkthrough method~\cite{herrmann2004socio,herrmann2009socio}. This method helps groups understand complex systems by discussing a diagram of the system step-by-step. We held the workshop twice with a total of 15 participants and made changes between sessions based on feedback. The workshop had four main goals: to help team members get to know each other, to align everyone’s understanding of the digital currency system we were designing, to refine the journey of a digital currency asset in our system, and to explore the human factors of the system from different users’ perspectives.

\begin{figure}
    \centering
    \includegraphics [width=0.7\textwidth]{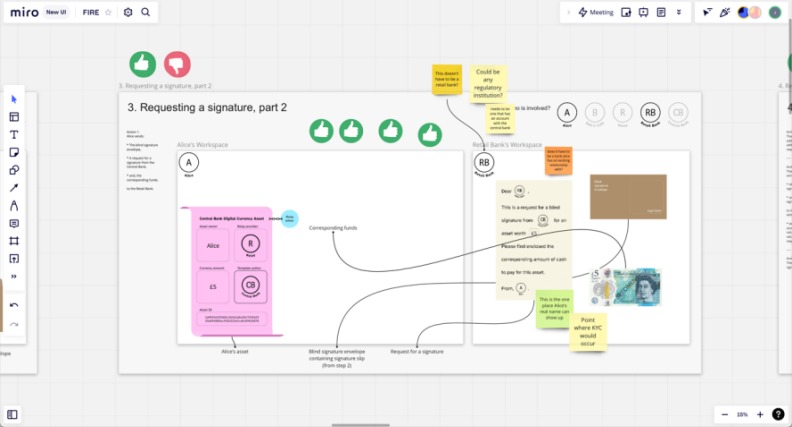}
    \caption{Blind signature request in iteration 1 of the internal workshop}
    \label{fig:enter-blind}
\end{figure}

Team members in this ``internal'' workshop included academics with backgrounds in computer science, finance, and design; representatives from different groups within a central bank; and representatives from different groups within a global-scale technology infrastructure solutions provider.  In the workshop, we utilised Miro to turn our current work-in-progress which focussed on the technical requirements of our system into an interactive experience. Participants engaged within our proposed technical system by taking on the role of assigned stakeholder roles and followed the process of creating, validating, and spending a digital currency asset using interactive elements (see Figure~\ref{fig:enter-blind}).

After each step of the process, the group engaged in a discussion to reflect on their experiences. This discussion allowed participants to gain a deeper understanding of the technical system and build their knowledge and perspectives with each step. The discussion also aided in addressing the issue we had in presenting difficult-to-understand digital currency systems. Discussions focussed mainly on clarifying the mechanics of the process; critical reflection on the process; and the discussion of alternative approaches and perspectives. 

\begin{figure}
    \centering
    \includegraphics [width=0.7\textwidth]{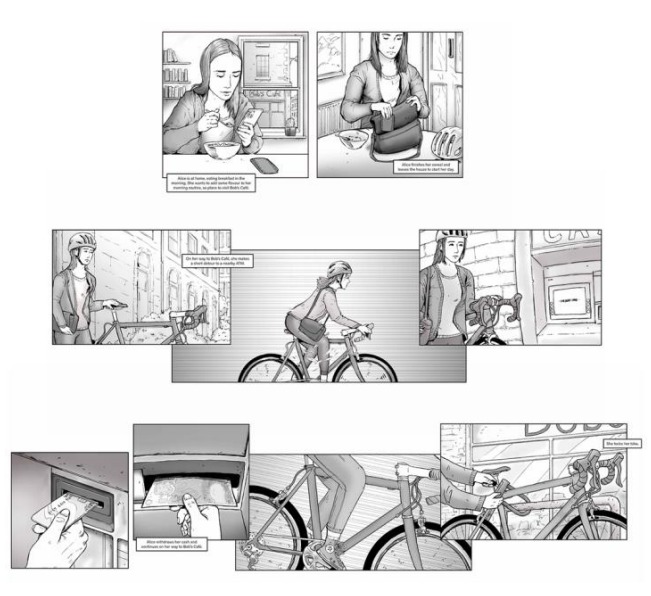}
    \caption{Problem framing comic (illustration by Peter Tilley)}
    \label{fig:enter-problem}
\end{figure}

This first workshop, however, revealed gaps, such as the need for a visual representation of how our system would be used in everyday scenarios. Visual narrative-driven tools, like comics, are powerful tools for facilitating communication and encouraging discussion, as they can convey complex ideas in an understandable medium. Researchers have effectively used comics for data visualization~\cite{10.1145/3290605.3300483}, code comprehension~\cite{10.1145/3411764.3445434}, and dissemination of qualitative findings~\cite{haughney2008using}. Through visual storytelling, comics allow complex concepts to be presented in a familiar format. This made the narrative-driven ability of comics to be a good solution to ground our technical system into a representational story that was potentially relatable for everyday people, such as buying a coffee at a shop (see Figure~\ref{fig:enter-problem}).

\begin{figure}
    \centering
    \includegraphics [width=0.7\textwidth]{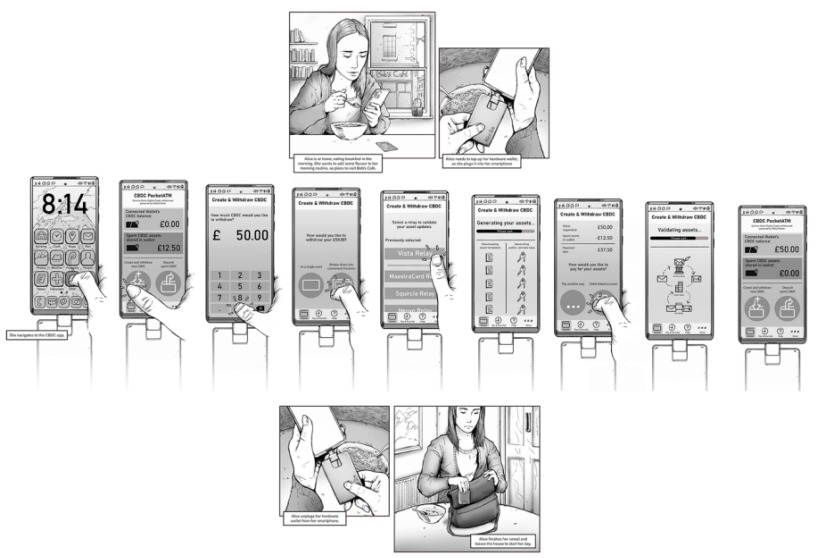}
    \caption{User journey comic (illustration by Peter Tilley)}
    \label{fig:enter-user}
\end{figure}

Applying the story-driven approach to identifying sticking points in the workshop directed the conversation towards human-centric issues, such as who would mint the digital currency tokens, how the role of a central bank would change in relation to our proposal, and what the benefits would be to vendors transacting within our digital currency system. These lines of enquiry were later used to refine the design and communication of our proposed digital currency system. The workshop was successful in achieving its objectives and enhancing participants’ comprehension of the digital currency asset lifecycle and the roles of stakeholders within it. The activity elicited questions and sparked discussions about the design of the system, and above all, it allowed us to iterate and deepen our understanding of both technical and user requirements within our proposed digital currency system, grounding participants in the user-centred and social-technical scenarios they might encounter themselves if they were to use such a digital currency system. The design team utilised the outcomes to refine the technical design of the non-custodial wallet as well as the system as a whole.

\subsection{Storytelling as a Method to Explore Digital Currency}

We continued with our goal to develop our digital currency system from a human-centred approach by extending our participant reach; we invited a total of 22 participants to another workshop titled ‘Stories of Central Bank Digital Currency.’ This workshop saw public sector, private sector, and academic organizations come together to engage in a multi-stakeholder evaluation of our digital currency system.  Participants included academic faculty members and students with research experience in computer science, mathematics, finance, management, and design; government experts from different groups within a central bank and financial regulators; and a variety of industry participants including global-scale financial technology infrastructure providers, a multi-national investment bank, technology consulting firms, digital currency firms, start-up technology businesses, standards development organisations, and well-known financial industry associations.  The main goal of this workshop was to gather feedback on our digital currency system proposal, with a specific focus on human factors and end-user journeys. We gathered user requirements for devices or interfaces that would allow end-users to interact with the system. As with the first set of workshops, an overall goal of identifying potential directions for the future development of the system was also central to this workshop, which was broken into two stages, ‘Mapping present-day transaction journeys’ and ‘Provotypes to elicit user insights and critique'.

Overall, the workshop centred on the theme of storytelling and narrative design. Storytelling more broadly has proven to be an essential tool in design, facilitating an understanding of how technology can support self-reflection around intense experiences such as grief~\cite{ataguba2018towards} and how stories can be conveyed through emerging technological mediums like virtual reality~\cite{ye2016storytelling}. Storytelling can also serve as a valuable instrument for comprehending and engaging individuals in existing socio-technical systems, like empowering youth to have a voice in the design of their environments~\cite{poplin2017engaging}. The workshops consisted of presentations and participatory activities that utilised storytelling to discuss the topics of cash, digital currency, and the ongoing work of the project.

\subsection{Mapping Present-day Transaction Journeys}

The term `user journey' is used to describe the steps a user takes when interacting with software, hardware, or any other product. It can be challenging to identify a user’s needs right away, but user journey mapping can help. This creative method, as described by Endmann and Keßner (2016), allows for a quick understanding of user processes and helps prioritise design concepts~\cite{endmann2016user}. Many researchers have used journey mapping to gain insight into a user’s experience in specific scenarios, such as using a library~\cite{marquez2015walking}. This method offers a 3-dimensional view of the user’s journey, providing more depth than other methods like personas~\cite{howard2014journey}.

\begin{figure}
    \centering
    \includegraphics [width=0.7\textwidth]{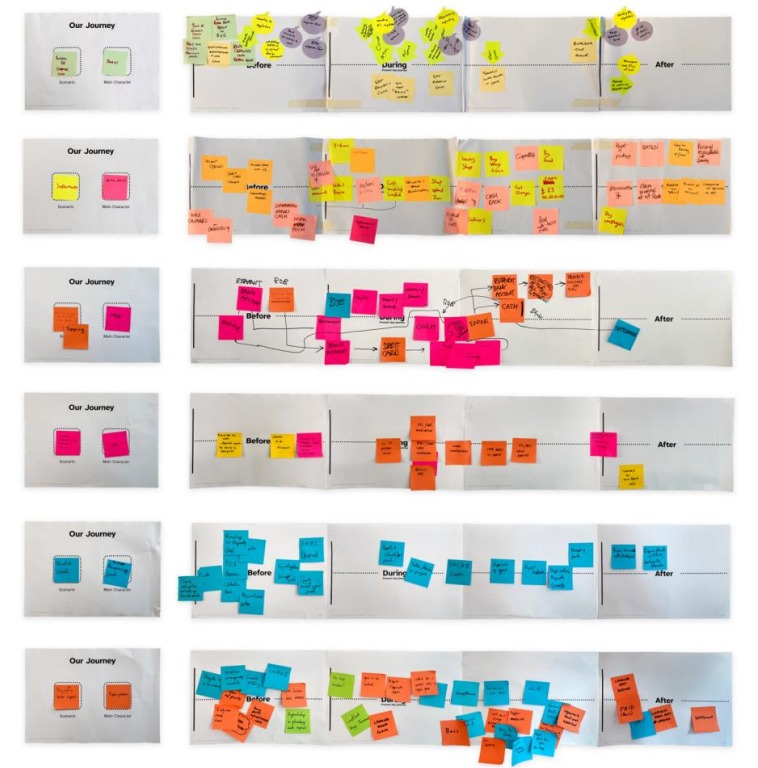}
    \caption{User Journey Mapping}
    \label{fig:mapping}
\end{figure}

We utilised the User Journey Mapping method by dividing participants into groups and guiding them through a step-by-step activity focussed on present-day payment scenarios. They created their User Journey Maps using Post-it notes(see Figure~\ref{fig:mapping}).

Each group then presented their journey map to the other participants. This step of the task would later prepare participants to critically evaluate our digital currency proposal by breaking down any pre-existing complexities of payment scenarios encountered in daily life. While creating their journey maps, participants exchanged knowledge with one another, which they could then use to formulate questions about current payment scenarios and our digital currency proposal. A total of six user journey maps were created by participants that explored current payment scenarios in various contexts: a bureau de change (group 1), a supermarket (group 2), a restaurant (group 3), an international transfer (group 4), receiving funds from a charity (group 5), and a boiler repair job (group 6).

The journey maps created by the participants illustrated the intricate ways in which payments are entangled with the social, political, economic, and, of course, personal circumstances of end-users. For example, group 2 presented a scenario in which a user at a supermarket splits their shopping basket into two forms of payment: certain goods to be purchased on their credit card and others, like cigarettes or lottery tickets, to be paid for with cash. The reason for this, as one participant explained, is ``because they don’t want their partner to know quite what’s going on.'' Group 1’s journey followed the various routes that foreign currency can take to end up in the register of a bureau de change, including the complex relationship between currency importation and international relations. Group 5’s journey considered several methods of donating to charity, including the use of automated tools such as round-ups on debit card transactions. At the end of this activity, participants were able to discuss complex user scenarios around payment options in different contexts. This provided valuable user insight and perspectives. This exploration bolstered our understanding of our digital currency proposal and the everyday decisions people must make around forms of money or ways to use their money.

\subsection{Provotypes to Illicit User Insights and Critique}

\begin{figure}
    \centering
    \includegraphics [width=0.9\textwidth]{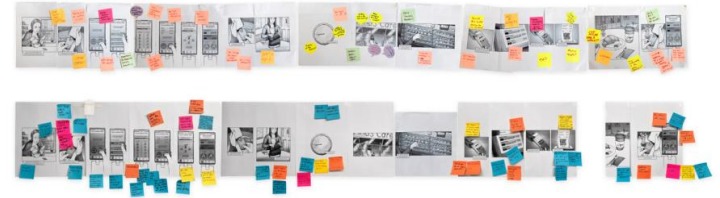}
    \caption{Comic strip of digital currency scenario with Post-it Notes}
    \label{fig:comic}
\end{figure}

After completing the user journey mapping task, participants embarked on the next stage of the workshop. They gathered around two ‘provotypes’---a portmanteau of `provocation' and `prototypes'~\cite{boer2012provotypes}---including a large-format print of the digital currency user journey comic and a 3D-printed model of the debit card-style digital currency hardware wallet featured in the comic. These tangible representations allowed participants to apply their insights from previous explorations of existing payment systems to our concepts. In smaller groups, they began a self-led discussion about the proposed system, capturing their conversation on Post-It notes (see Figure~\ref{fig:comic}) and applying them to relevant areas of Alice’s journey, the person in the comic.

This activity provided an opportunity for participants to offer feedback on both the technical aspects of the digital currency system design and the narrative approach used to convey it. Participants were encouraged to apply their insights from the first activity to the future user journey presented to them, considering how the social aspects of the digital currency system might impact its use.

Drawing on insights from this workshop, we began to understand some additional considerations and interesting situations for which our system might need to adapt. Participants considered digital currency as a type of payment to be used within given scenarios, such as for low- to medium-value everyday purchases or to enable government uplift payments to people without bank accounts. Trust was also a key consideration, with questions raised about how a central bank could sign off on the creation of anonymised tokens and why a user would trust their hardware wallet or the app they use to make transactions. Complex notions of inclusion and exclusion were also discussed, with concerns that the highly technical nature of the system could alienate people already excluded from digital services. Conversely, it could also allow those without bank accounts to make and receive digital payments, expanding concepts of financial inclusion. Other important considerations included the system’s capacity to function offline or on local networks, the visual appeal and user-friendliness of the hardware wallets, and the measures in place to mitigate loss in the event that a hardware wallet is lost or stolen.

We acknowledge that we cannot simply assume that users will be able to migrate seamlessly from mobile banking apps popular today to a new paradigm of non-custodial wallets.  Even if users were to appreciate the security implications of not being able to rely upon a custodian to bail out their losses, and either carry a separate device for digital currency or accept responsibility for risks associated with having a digital currency app on a smartphone, some differences in the user journey between mobile banking apps and digital currency would remain.  For example, unlike a mobile banking app, a digital currency device or app would not require identity-based authorisation when making a payment, because the assets are held directly on the device.  Also unlike a mobile banking app, users of a digital currency device or app would be required to load it with digital currency from time to time, either over the Internet or via a kiosk.  Finally, the mechanism for resolving problems with botched purchases or stolen assets would be different.  Rather than relying upon a custodian to adjudicate claims, users of a digital currency device or app would have a different process, perhaps involving police or insurance firms, to demonstrate that they had legitimately owned an asset that had gone missing or had fallen into the wrong hands.  All of these differences follow from the fact that unlike bank deposit assets, which are claims on the balance sheet of a custodian, assets that are directly possessed and controlled by their users are not a matter of credit but a matter of direct ownership.

As a whole, the workshops provided valuable feedback and user insights that helped shape the design of our digital currency system. Key outcomes were identified from running these user-centric workshops. Firstly, it became clear that clearer definitions were needed for how our digital currency would integrate into the UK economy. This included specific definitions of the types of transactions for which the digital currency system is best suited and considerations around the types of public and private organisations responsible for designing, building, and maintaining the physical and digital infrastructure required to run the system. Lastly, clarity was needed in our framing of the benefits and drawbacks of the system. Communication is essential and must be understandable by both experts and non-experts. Some users just want to pay and move on with their day without considering technical specifications. However, if they do want to understand the privacy of the system and build trust that it will work as advertised, then this information must be accessible to all levels of users. Ultimately, we are designing a socio-technological output; therefore, framing will be key for people to want to use the system.

We understand that there are existing non-custodial wallet hardware devices, such as the Nano X developed by Ledger SAS, that could be used to store digital assets such as the ones that we propose.  Our purpose of exploring new designs is not to validate or invalidate such existing devices, but to establish baseline requirements and understand the trade-offs between different potential designs.  Ultimately, we understand that the device must be able to store data, run some cryptographic operations, provide for end-user interaction, and communicate with other devices, but the potential design space is large.  We also understand that for public infrastructure, such as a CBDC or an institutionally supported reserve-backed token or stablecoin solution, we would need to ensure that the solution is low-cost and accessible to a broad population, including marginalised groups.  While it might be possible that solutions such as the Nano X would be appropriate and sufficient for this purpose with relatively few modifications, we seek to comprehend the design space as a whole.

Our user-focussed approach to digital currency design has emphasized the importance of considering the user’s perspective. By using provotypes to integrate the technical aspects of the digital currency system into a narrative-driven story, we have been able to elicit deep and useful design insights that can shape the technical requirements of our system. This approach has also helped us identify flaws in our framing, leading us to iterate on our workshop approaches and continue refining the technical requirements for our proposed digital currency system.

\subsection{Towards creating a non-custodial wallet}

After gathering valuable insights from our users on what should inform our digital currency system, we are now faced with the task of delivering this system. To inspire our digital currency system design, we once again turn to cash, this time by looking at the concept of a wallet. Wallets carry cash and debit in the form of plastic and have become electronic, having been associated with the notion of money for decades. Therefore, to begin our exploration of how a user will come to use their digital currency, we examine the two types of these wallets: custodial and non-custodial. Custodial wallets are a range of digital escrow wallets that store crypto assets externally, outside of a user's device. Non-custodial wallets, on the other hand, can be either software-based or hardware-based and can satisfy different preferences related to security and privacy~\cite{gomzin2022choose}. 

However, custodial wallets are not easy to use and often require a level of knowledge. Their user design is typically oriented towards experts, with a complex registration process and a high potential for errors when paying in cryptocurrencies~\cite{frohlich2021don}. Both experts and non-experts can face financial loss due to the relative lack of user-friendliness of custodial wallets, which could be mitigated by better mimicking traditional banking methods and providing pre-education on their use~\cite{voskobojnikov2021u}. Designing a user-friendly experience for novel concepts like digital currency or crypto requires balancing functionality with user preferences such as security and risk minimisation, without overwhelming the user with technical details.

Just as cash represents trust, a digital currency must also instil trust in its users. Traditional payments are backed by institutional protections that help establish trust, while decentralised crypto assets lack these protections, making trust harder to establish~\cite{borio2019money}.  As such, the design of a digital currency wallet must provide uncompromising security, privacy, and identity protection while also supporting the regulations and protections similar to those that apply to centralised payment methods. This would allow for trust and give users equal options to choose their preferred form of payment.  Achieving privacy by design requires a transaction paradigm that differs substantially from the four-point pull payments that are characteristic of retail transactions in much of the world.  Specifically, privacy requires a mechanism for unlinking the identities of the transacting parties from each other in the transaction channel.  This can be achieved with blind signatures, as described earlier, or via zero-knowledge proofs, as are used in some privacy-enabling cryptocurrency systems, such as Zcash and Monero.  In addition, since one of the parties must communicate with a third party to achieve fair exchange, we suggest that in most cases that party should be the payee rather than the payer.  As a merchant, the payee is more likely to be subject to tax obligations and business regulations than the payer and, as a consumer, an ordinary individual is more likely to suffer harm through the profiling of his or her transactions.

The requirements underpinning our design assume that the payer would be anonymous and the payee would be known, with blind signatures ensuring that the payer's identity is not associated with the transaction~\cite{goodell2021scalable}.  Following the logic articulated by Auer and Böhme, we understand that privacy by design requires the system to use tokens rather than balances, as account balances require the creation of an identity which is accessed on successive transactions~\cite{auer2020a}.  We also assume that compliance requirements (such as tax and sanctions policy) would be enforced on the side of the payee, who receives the digital assets.

A person using digital currency should not need service relationships to use their money. For this reason, we turned to the non-custodial design, which provides hardware and software options for wallet design and allows for better implementation of user-friendliness in its core design. Our design concerns follow core principles: the wallet must store digital currency information, perform cryptographic functions, and have the ability to send and receive data.

\subsection{Wallet designs}

To ensure privacy and security, wallets must not be identifiable and must not require registration or trusted computing. We considered several wallet design options and approaches that can satisfy our requirements to varying degrees. These were represented by four concept wallets:

\begin{itemize}

\item The \textit{``I can’t believe it’s not a phone''} wallet: a lightweight independent phone-like device with no wireless capability that runs a light OS on top of the core cryptographic functions required of the wallet (see Figure~\ref{fig:notphone}).  Its full-featured design means that it can simplify and expedite sophisticated operations, such as token selection and management, and it has a customisable, fasionable UI.  Disadvantages include the price (additional display and computation performance requirements mean that it will be considerably more expensive than alternatives, possibly impeding its usefulness as a publicly distributed device) and the fact that it is almost certain to require external power or charging.  Also, from a security perspective, the complex OS means that it has a potentially larger attack surface.

\begin{figure} [H]
    \centering
    \includegraphics [width=0.4\textwidth]{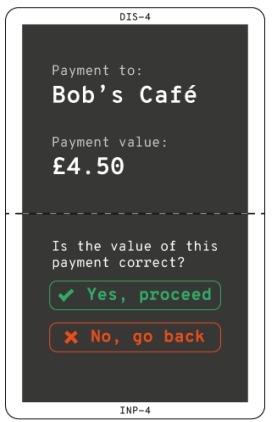}
    \caption{The ``I can’t believe it’s not a phone'' wallet.}
    \label{fig:notphone}
\end{figure}

\item The \textit{``card plus''} wallet: a device with no user interface and must be connected to a companion device such as a phone (see Figure~\ref{fig:cardplus}).  The lack of a user interface means that it requires a companion device, which the user must trust.  It might also require more steps of interaction to communicate between the companion device and the interface provided by the transaction counterparty.  Advantages of this wallet design include the fact that its thinness means that it can be carried in an enclosure, as if it were a card, as well as the fact that its lack of external buttons or user interfaces means that it can rely upon the accessibility features of a companion device.

\begin{figure}[H]
    \centering
    \includegraphics[width=0.4\textwidth]{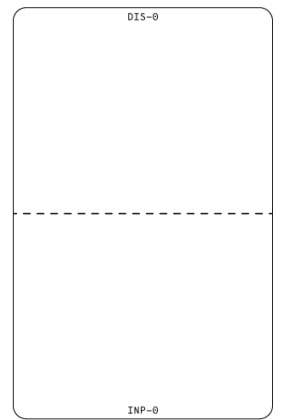}
    \caption{The ``card plus'' wallet.}
    \label{fig:cardplus}
\end{figure}

\item The \textit{``final check''} wallet: a mix of the two aforementioned wallets, with a simple numerical user interfaces that allows the user to verify the transaction, as it can show transaction information and can be physically plugged into other devices like a point-of-sale machine (see Figure~\ref{fig:final}).  This wallet offers better security than solutions without a user interface, since users can validate for themselves, using hardware that they possess and control, the amount of money that they are authorising to pay in a transaction.  Like the ``card plus'' wallet, it can be designed to be thin, with minimal power consumption.  To achieve these benefits, however, user interaction is limited to a value check and final authorisation only.

\begin{figure}[H]
    \centering
    \includegraphics [width=0.3\textwidth]{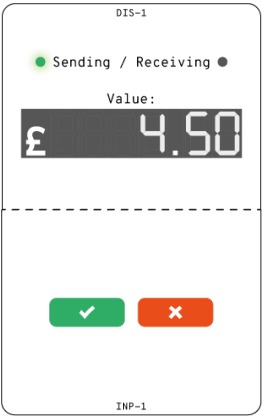}
    \caption{The ``final check'' wallet.}
    \label{fig:final}
\end{figure}

\item The \textit{``minimum standalone''} wallet: a direct upgrade of the ``final check'' wallet, although with a more adept user interface and input, numerical pad, and the ability to verify transactions through buttons and visual representations of transaction data information (see Figure~\ref{fig:standalone}).  With this wallet design, payers can make choices about the amount of the payment, select specific tokens, and manage their token inventories.  Payers may be able to verify the identity of the recipient, thus offering an additional layer of security verification, and they may be able to protect their devices with a PIN code as well.  However, the somewhat larger UI and display might require some additional thickness and might consume more power, to the extent that solar panels or an external power source (e.g. necessitating periodic charging) might be necessary.

\begin{figure}[H]
    \centering
    \includegraphics [width=0.3\textwidth]{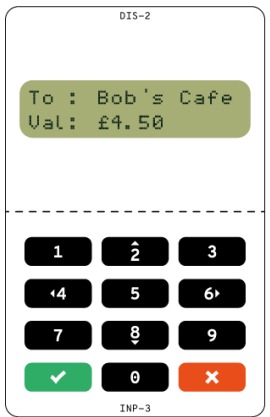}
    \caption{The ``minimum standalone'' wallet.}
    \label{fig:standalone}
\end{figure}

\end{itemize}

The development of a collection of non-custodial wallets provided us with new insights into the technical and hardware requirements that aimed to balance ease of use, transactional security, convenience, and affordability. Using a non-custodial wallet, especially a hardware version, incurs an automatic cost associated with acquiring the physical hardware. That is why the presence or absence of component parts, such as an LCD screen, can make a big difference in the cost ratio, with a concomitant impact on user experience.

\begin{figure}[H]
    \centering
    \includegraphics[width=0.7\textwidth]{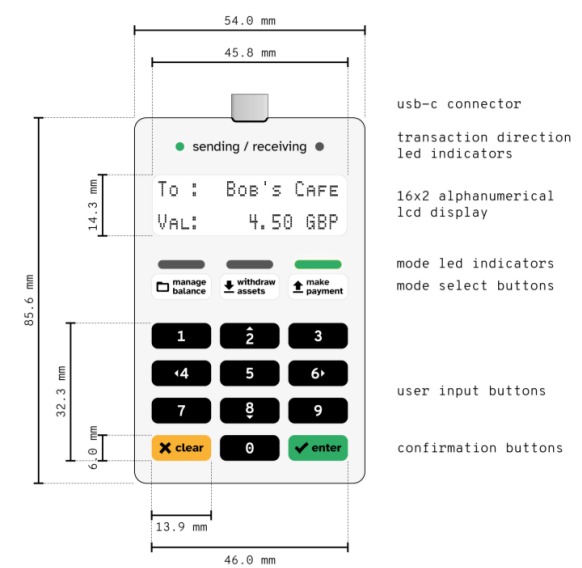}
    \caption{Features of the minimum standalone wallet.}
    \label{fig:standalone-2}
\end{figure}

\begin{figure}[H]
    \centering
    \includegraphics[width=0.6\textwidth]{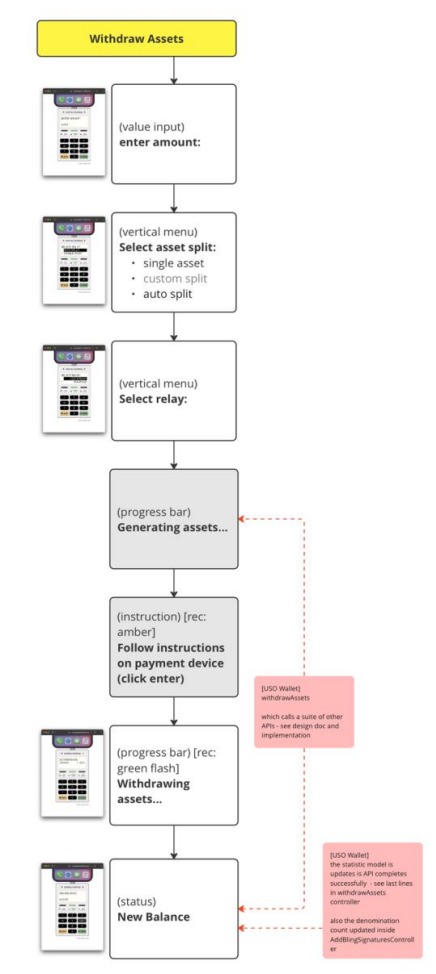}
    \caption{Schematic representation of the process flow for withdrawing assets.}
    \label{fig:withdrawing}
\end{figure}

\begin{figure}[H]
    \centering
    \includegraphics[width=0.8\textwidth]{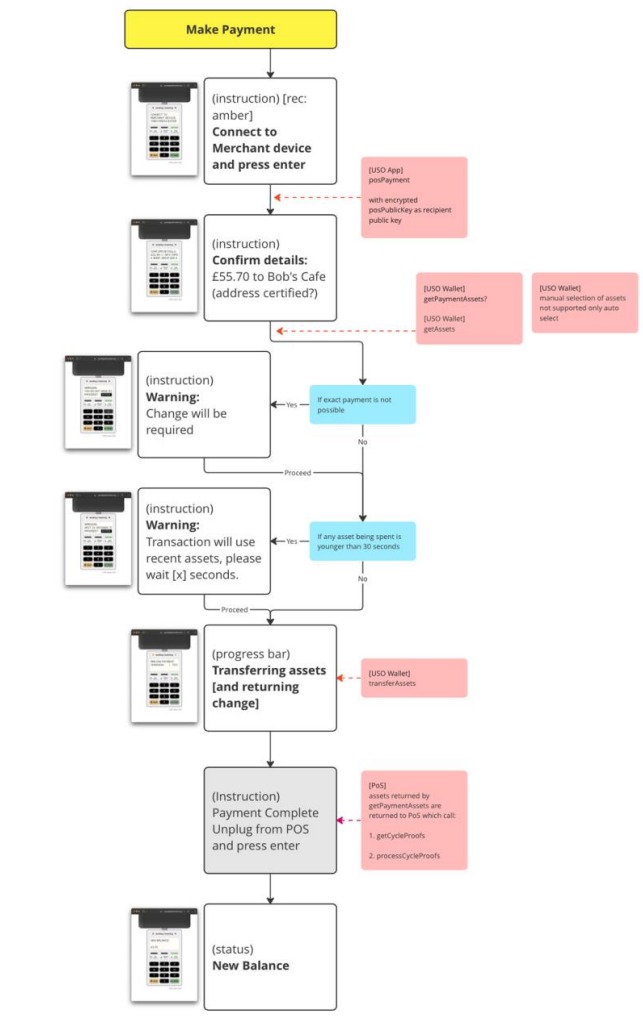}
    \caption{Schematic representation of the process flow for making a payment.}
    \label{fig:payment}
\end{figure}

Experimentation with these design requirements showed that \textbf{The minimum standalone wallet} (see Figure~\ref{fig:standalone-2}) was the closest match to our requirements that we could find.  Although it is more costly than the ``card plus'' wallet, its security properties and user-friendliness made for an overall better non-custodial wallet, thus providing a suitable proof of concept. However, this choice should not be seen as limiting the potential design space for wallets. We firmly believe that a variety of viable wallet designs is necessary to meet the diverse needs of users in different payment scenarios. A simulation of the wallet was created using the \texttt{Next.js} frontend JavaScript framework, wherein we were able to simulate the functionality required to view the contents of the wallet, withdraw assets (see Figure~\ref{fig:withdrawing}), and make a payment (see Figure~\ref{fig:payment}).

\subsection{Observations}

In addition to helping us to understand the trade-offs between our wallet designs, our study participants also provided other important points of feedback as well, specifically including:

\begin{itemize}

\item\textit{Making change.}  Study participants noted that in a token-based system with denominations like the one we propose, customers will often not have a set of tokens that sum to exactly the right amount for a given transaction.  Thus, we will need a mechanism for making change in the absence of an identifying relationship between a customer and a service provider, such as a bank.  For this purpose, we proposed a solution.  Assuming that the merchant has access to a financial institution that can facilitate the minting of tokens, the customer can submit a payment that includes overpayment along with a blinded request for a set of tokens whose total matches the size of the overpayment.  The merchant can then relay this blinded request along with the overpayment itself, so that the new tokens can be issued and returned to the customer.  Because this solution requires the merchant to have access to minting infrastructure, it introduces an additional constraint beyond the general case, wherein a customer can transfer assets to a merchant as long as there is a mutually trusted third party to witness the transaction.  However, this requirement for connectivity does not exceed the requirements of existing digital payments infrastructure at the point of sale.

\item\textit{Loading tokens to the wallet device.}  Study participants noted that loading tokens to the wallet device, in advance with sufficient time delay, represents a new step in the transaction process that may be new to users of mobile banking apps.  It is not new to users of cash who regularly visit a bank or ATM to withdraw cash for making transactions, and users of cash might be gratified to learn that with a digital currency system like ours, the risk of the bank or ATM requiring a fresh delivery of cash is mitigated, and it is also possible for users to avoid the trip to the bank or ATM entirely by using a mobile banking app or web service to establish a connection to a customer's financial institution, which can then be used to load tokens to the digital currency wallet device.  However, even this streamlined process might seem cumbersome for someone accustomed to a one-step transfer from his or her bank directly to the merchant's bank, because it requires receiving digital assets in advance of the transaction, with a time gap.  It is certainly possible to connect digital wallet devices to the Internet, perhaps via a companion device such as an individual's mobile phone or personal computer, and script the authorisation of withdrawals at scheduled or random times, for example, when a user is sleeping.  The design allows for a multitude of potential solutions, such that it might be inappropriate to opine about which solution is best-suited for a particular use case.

\item\textit{Whether to pay a merchant or the merchant's bank.}  Study participants asked whether the merchant or the merchant's bank would receive assets from the payer.  We had deliberately left this detail unspecified.  Although we designed the system so that merchants could hold assets directly, we also noted that regulators would most likely require identifying information about the payee, such as a bank account number, to be included by the payer in the transfer.  It might be reasonable to imagine that it might be convenient for the payee's bank to intermediate the transaction on behalf of the payee in many circumstances, and doing so would make the act of receiving money more similar to the process of receiving money via card payments or interbank payments today.  However, because it is also possible for a consumer to pay a merchant, even if regulators require the merchant to be identified, without the involvement of a custodian on either side of the transaction, it might be important to support this path as a way to encourage competition among service providers.

\item\textit{Fully offline payments.} Our study participants did not express much concern about supporting payments in which both parties are offline without network access to a mutually trusted third party.  This supports our argument that such payments constitute a decreasingly important use case in general, and that there will always be a multiplicity of methods for making payment, rather than one payment system to rule them all.  We note that since fair exchange requires a trusted third party, alternatives to a networked third party might include a third party embedded into a device (trusted computing) or deliberate tolerance for double-spending in limited situations.  Rather than require or recommend such approaches, which would introduce additional trust requirements for users of money, we have instead chosen to focus on use cases in which such approaches are not required.  Note that the requirement for one party to have network connectivity to a mutually trusted third party is a lesser requirement than what is needed for making change.

\end{itemize}

In summary, our proposed digital currency system is designed to exist within a diverse ecosystem of payment scenarios, stakeholders, and worldviews. We have focussed on over-the-counter payments as our core use case, but we plan to explore additional scenarios such as online payments, peer-to-peer transactions, and high-value payments. Our next step is to have our non-custodial wallet working in a simulated back-end environment, to research further use cases and implications of our proposed system. While we believe that a non-custodial hardware wallet is an important option for digital currency users, we also anticipate the development of alternative wallet solutions by market actors. These may include smartphone applications and account-based services, which may offer greater convenience at the expense of privacy and control. It is important for users to be aware of these trade-offs when choosing the right wallets for their intended use cases.


\section{Limitations and future work}

Our approach to the digital currency non-custodial wallet is just one of many possible approaches. We have yet to test our final prototype with users to confirm that it meets our wallet requirements in real-world scenarios with diverse users. Our prototype was built using off-the-shelf hardware, such as the Pinephone.\footnote{https://www.pine64.org/} However, we suspect that future non-custodial hardware will be custom-made, potentially featuring variations in factors such as affordability and security.

Despite our thorough and user-centric approach, there is still room for improvement. For instance, co-design methods could be used in the decision-making process for our proof-of-concept wallet. Of course, these are limitations based on our research parameters.

Future work could take a truly bottom-up approach all the way to testing and deployment.  In particular, we envision developing provotypes for the various designs and deploy them in a test environment with real users, perhaps as a prelude to a pilot in a closed e-money environment, such as a university refectory or concessions network.

Other approaches may include smartphone applications that allow users to store digital currency tokens locally or account-based services that allow storage on the cloud. However, these approaches represent a shift from our privacy-by-design paradigm to a privacy-by-trust paradigm. Users must be aware of the trade-off between security and convenience.

As digital currency comes to fruition, what we currently deem as non-secure may advance and change. Designing for digital currency is currently based on proof-of-concept implementations and working within a present space of knowledge. Some future variables and trade-offs are still unknown, making it a fertile design space, albeit one with many limitations.

Other questions persist around the limitations in our design, such as the question of what happens if the device is stolen or lost. Although we have defined some proposed approaches for handing such scenarios, such as backups, selective disclosure of private information, and insurance services, these are beyond the scope of this article.  Another important consideration is: Will merchants accept our proposal? These are questions that still remain and require future work to answer, and we anticipate that engaging more users and merchants will uncover answers to these questions.

Overall, we encourage designers to pursue not only the challenge of engaging with non-custodial wallets, but also other aspects of delivering digital currency and the systems that will ultimately come to underpin them.

\section{Design Takeaways}

Cash provided a crucial starting point for designing a digital currency system, as there was no pre-existing method or framework to determine the necessary requirements for such a system. While there are many prospective architectures for digital currency systems that can be produced, each with its own unique set of methods and approaches, we believe that the insights we gained from our exploration could be valuable for others looking to enter this space. We provide basic guidelines for designers to follow or adapt, as part of the process of establishing a starting point for their own digital currency systems:

\begin{itemize}

\item Use current existing concepts of money, especially cash, as a starting point for designing a digital currency system, considering its established values and practices.  Since there was no pre-existing method or framework for designing a digital currency system, the insights gained from exploring cash proved to be invaluable in determining the necessary requirements for our own system.

\item Incorporate the opinions of diverse stakeholders, both within and outside the project team, to ensure an inclusive design process. Understand that digital currency discourse is heavily oriented towards expert knowledge. Use narrative-driven tools like comics or alternative methods to convey complex digital currency or payment information to engage diverse expert and non-expert stakeholders. Always aspire to establish an inclusive design process, to ensure that more people can engage with confidence despite limited knowledge.

\item Help people envision how they currently use money and how digital currency could fit into their existing practices and preferences.

\item Design digital currency systems to work in tandem with other forms of money and services, giving users flexible autonomy in what they purchase, how they pay, and whether such payments should offer anonymity to payers or not.

\item Acknowledge that introducing new options can result in some existing options being taken away.  To ensure flexibility, a designer might look to balance digital currency system design by instead focusing design efforts towards keeping established forms of money from disappearing, to uphold the diversity of payment options.

\item Designers might take an international approach to designing a digital currency system to ensure that it is compatible with diverse monetary practices and preferences.

\item Design digital currency systems to be inclusive and accessible by considering diverse options for accessing digital currency assets, for instance allowing digital deposits without the need for a bank account and balancing affordability, design, user experience, accessible hardware, and infrastructure whenever possible. Explore digital currency from both inclusion and exclusion angles to inspire novel technologies, services, infrastructures, or even tangible methods for accessing digital currency.

\end{itemize}

\section{Conclusion}

In this paper, we explored the design of a non-custodial wallet, a device that enables users to store and spend digital currency, including but not limited to central bank digital currency (CBDC), in various payment scenarios. We drew on established values and practices of current forms of money, such as cash, to inform our design. We incorporated the opinions of diverse stakeholders, both within and outside the project team, to ensure an inclusive design process. We used narrative-driven tools, such as storytelling and metaphors, to make digital currency more accessible and comprehensible for users. We also elicited user feedback and critique on our digital currency system proposal by using provotypes. Our research revealed some basic guidelines for designing digital currency systems, such as designing for compatibility with other forms of money, ensuring accessibility and inclusion, and balancing the technical and social aspects of digital currency. We also highlighted the importance of protecting established forms of money like cash, as a way to maintain flexible payment options and prevent their decline in usage. We demonstrated the innovative potential of digital currency system design to protect the privacy and security of users while ensuring user-friendliness and giving more people more choices in their payment options. We encourage other designers to explore this novel opportunity to critically consider the design of money, which has the potential to shape everyday life.

\section*{Acknowledgements}

The authors acknowledge the support of the Stellar Development Foundation and
the Systemic Risk Centre at the London School of Economics.  The authors also
acknowledge EPSRC and the PETRAS Research Centre EP/S035362/1 for the FIRE
Project.  Finally, the authors thank anonymous reviewers for their useful
comments.

\bibliographystyle{plain}
\bibliography{bib.bib}

\end{document}